# ELECTRON COOLING PERFORMANCE AT IMP FACILITY*

Xiaodong Yang[#], Jie Li, Lijun Mao, Guohong Li, Xiaoming Ma, Tailai Yan,
Mingtao Song, Youjin Yuan, Jiancheng Yang, Ruishi Mao, Tiecheng Zhao, Peng Li,
Wei Zhang, Dayu Yin, Weiping Chai, Huan Jia, Wenheng Zheng, Xiaohu Zhang,
Institute of Modern Physics, CAS, Lanzhou, 730000, P. R. China.

*Abstract*

The ion beam of $^{58}Ni^{19+}$ with the energy of 6.39MeV/u was accumulated in the main ring of HIRFL-CSR with the help of electron cooling. The related angle between ion and electron beams in the horizontal and vertical planes was intentionally created by the steering coils in the cooling section after maximized the accumulated ion beam in the ring. The radial electron intensity distribution was changed by the ratio of potentials of grid electrode and anode of the electron gun, the different electron beam profiles were formed from solid to hollow in the experiments. In these conditions, the maximum accumulated ion beam intensity in the 10 seconds was measured, the lifetime of ion beam was measured, simultaneously the momentum spread of the ion beam varying with particle number was measured during the ion beam decay, furthermore, and the power coefficient was derived from these data. In additional, the momentum spread in the case of constant particle number was plotted with the angle and electron beam profile. The oscillation and shift of the central frequency of the ion beam were observed during the experiments. The upgrade and improvement in the CSRm cooler and the progress in the CSRe cooler were presented. These results were useful to attempt the crystal beam forming investigation in the CSR.

## MAIN WORKS IN CSR

- $^{209}Bi^{36+}$ Accumulation and Acceleration in CSRm
- Experiments related to cancer therapy [1]
- Patients treatment
- Mass measurement [2]
- Prophase Experiments on recombination [3]

### Accumulation and acceleration of $^{209}Bi^{36+}$

A new superconducting ECR ion source SECRAL developed by IMP has started operation to provide high intensity heavier ion beam. $^{209}Bi^{36+}$ delivered by the SECRAL was accelerated by smaller cyclotron SFC to 1.877 MeV/u and then injected into CSRm. The average pulse intensity was about 1.8 μA in the injection line. The average pulse particle number of $^{209}Bi^{36+}$ was about $7.3 \times 10^6$ in one standard multi-turn injection. With the help of electron cooling of partially hollow electron beam, $4.4 \times 10^7$ particles were accumulated in the ring after 67 times injection in 10 seconds, and $1.3 \times 10^7$ particles were accelerated to the final energy of 170 MeV/u. The DCCT signal of $^{209}Bi^{36+}$ beam was displayed in Fig. 1 during accumulation and acceleration with the help of electron cooling.

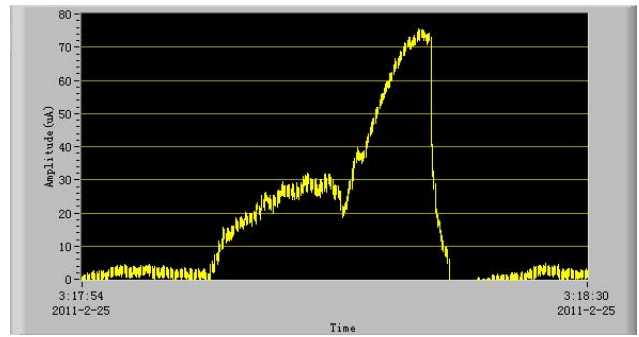

Figure 1: DCCT signal of $^{209}Bi^{36+}$ beam accumulation and acceleration with the help of electron cooling, 1.87MeV/u---170MeV/u.

## BEAM ACCUMULATION IN CSR

The main function of electron cooler in CSRm was heavy ion beam accumulation. The accumulation efficiency was related with a lot of parameters of storage ring and electron cooler, such as the work-point setting, closed-orbit, electron density, and angle between electron beam and ion beam. At the beginning, the electron beam alignment was done to maximize the accumulated ion beam intensity. This setting was defined as "0" angle. The related angle between ion and electron beams in the horizontal and vertical planes was intentionally created by the steering coils in the cooling section after maximized the accumulated ion beam in the ring in the case of fixed storage ring parameters setting and electron beam parameters. After ion beam accumulation, the ion beam intensity and longitudinal signal were recorded by the DCCT monitor and Schottky probe during the ion beam decayed, and the maximal accumulated ion beam intensity in the 10 seconds interval was derived from the DCCT signal. The dependence of the ion beam intensity on the related horizontal and vertical angle was presented in the Fig. 2a and 2b. The angle between ion and electron beams reflected the temperature of electron in the system. In the case of the fixed electron beam current, ion encountered different electron temperature. The temperature influenced the cooling force and cooling time. The cooling force varying as the angle was reported in the reference [4]. The cooling force approached to maximal value in the perfect alignment between ion and electron beams, and the cooling time was minimal. One can see the maximal accumulated ion beam intensity was obtained near the zero angles in both panes. Another main parameters was the electron beam profile, the radial electron intensity distribution was changed by the ratio of potentials of grid electrode and anode of the electron gun, in this sense, and the different electron beam profiles were formed from solid to hollow in the experiments. In these conditions, the maximum accumulated ion beam intensity in the 10 seconds was plotted in Fig. 2c. One can find that the maximal ion intensity attained near Ugrid/Uanode=0.2, this results was similar as the previous results from CSR [4]. This indicated that the accumulation efficiency approached maximal in the partially hollow electron beam.

___________________________________________
*Work supported by The National Natural Science Foundation of China, NSFC(Grant No. 10975166, 10905083, 10921504)
[#]yangxd@impcas.ac.cn

## EXPONENT

The scaling law of momentum spread varying as the stored particle number was reported in many references [5, 6, 7]. A lot of result was presented from the aspects of theoretical and experimental, in additional, some simulated results was presented. The power coefficient ξ was derived from the recorded iqt data by the Schottky probe in this experiment. The results were presented in the Fig. 3a and 3b. From the diagram one can see that the power coefficient became bigger in the case of a bigger related angle between ion and electron beams. It was near 0.3 in the smaller angles. These were consistent with the results of the references [7, 8]. When the electron beam was solid, the power coefficient was bigger. For a hollow electron beam it became smaller. The power coefficient as a function of the electron profile was shown in the Fig.3c.

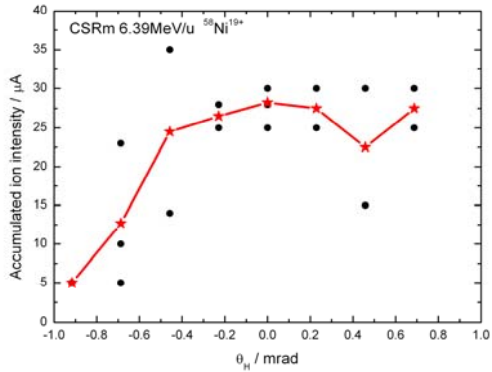

Figure 2a: The dependence of accumulated ion intensity in 10s on the related horizontal angle between ion and electron beams.

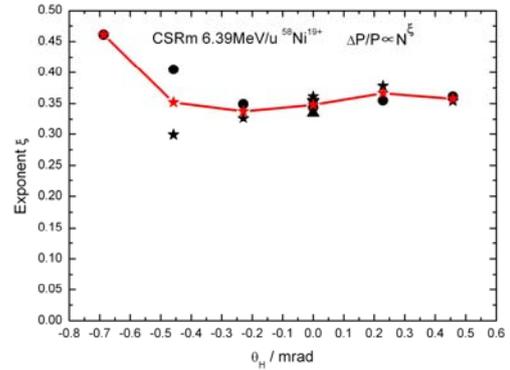

Figure 3a: The power coefficient varies with the related horizontal angle between ion and electron beams. The momentum spread was scaled with the particle number as power law $\Delta P/P \propto N^\xi$. The exponent in the figure is the power coefficient $\xi$.

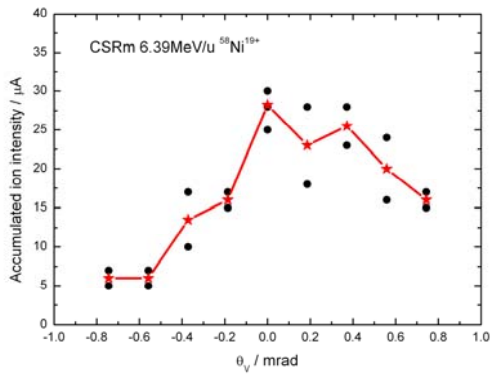

Figure 2b: The dependence of accumulated ion intensity in 10s on the related vertical angle between ion and electron beams.

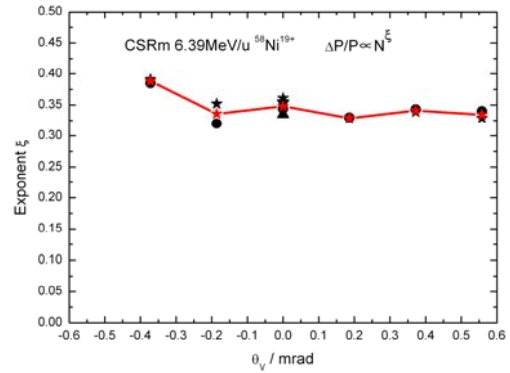

Figure 3b: The power coefficient varies with the related vertical angle between ion and electron beams.

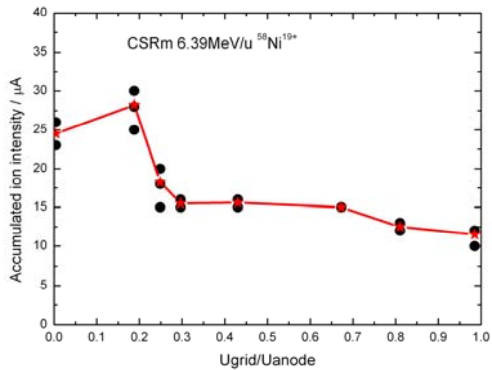

Figure 2c: The dependence of accumulated ion intensity in 10s on the profile of electron beam.

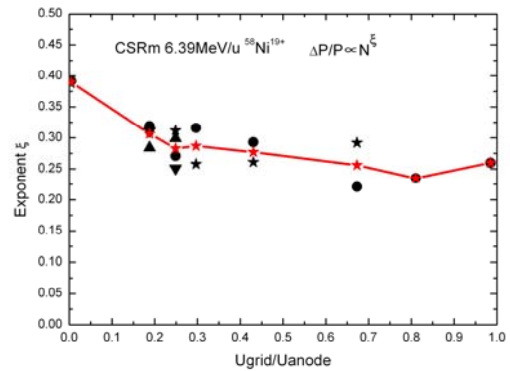

Figure 3c: The power coefficient varies with the profile of electron beam.

## MOMENTUM SPREAD

The momentum spread of the ion beam after electron cooling was an important parameter in the storage ring. It indicated the minimal value of the ion temperature after the equilibrium. The momentum spread versus the related horizontal and vertical angle between ion and electron beams was illustrated in the Fig. 4a and 4b in the circumstances of fixed particle number. The momentum spread was smaller in the condition of bigger angles, whereas the momentum spread was bigger under the circumstances of smaller angle. These results were good agreement with the reference [5, 7]. In the Fig. 4c, the momentum spread versus the electron beam profile was shown. These results revealed that the momentum spread was smaller in the condition of solid electron beam, and the bigger momentum spread in the case of hollow electron beam. For the same electron current the space potential did not change outside the electron beam, but the potential continuously increased inside for the solid electron beam. The potential was the reason of variation of the ion beam momentum spread with the changing of electron beam profile. When the ratio of Ugrid/Uanode increased, the profile of the electron beam was close to the hollow beam, the suppression of the electron beam energy decreased.

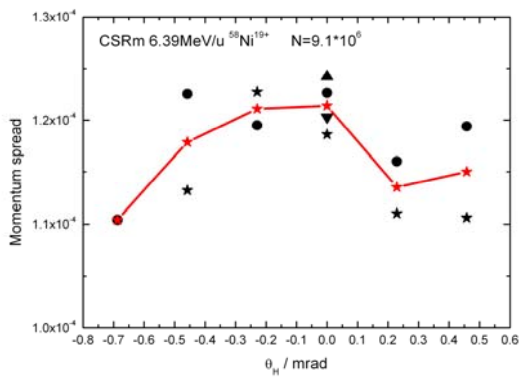

Figure 4a: The momentum spread versus the related horizontal angle between ion and electron beams in the case of fixed particle number $N=9.1*10^6$.

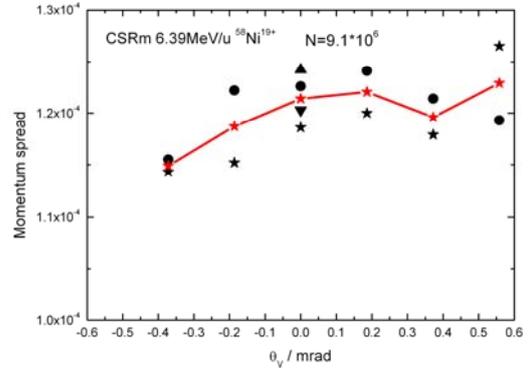

Figure 4b: The momentum spread versus the related vertical angle between ion and electron beams in the case of fixed particle number $N=9.1*10^6$.

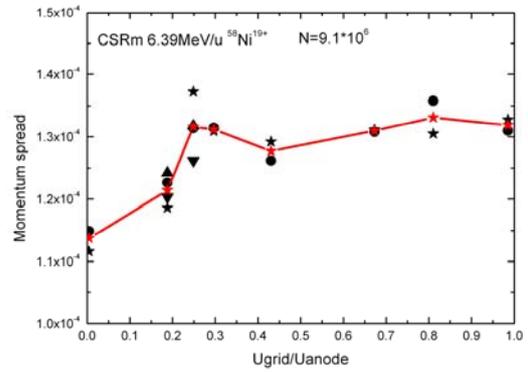

Figure 4c: The momentum spread versus the profile of electron beam in the case of fixed particle number $N=9.1*10^6$.

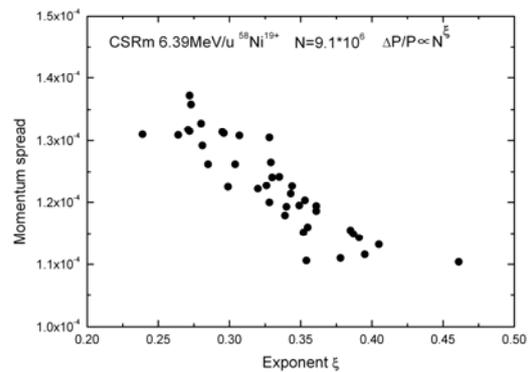

Figure 5: The momentum spread as a function of the exponent (power coefficient) in the case of fixed particle number $N=9.1*10^6$.

In the Fig. 5, the momentum spread as a function of the exponent (power coefficient ξ) was plotted. This result was similar as the reference [7]. The smaller momentum spread emerged in the case of a bigger exponent.

# LIFETIME

The lifetime of stored ion beam was a complicated problem in the storage ring, the factors influenced the lifetime included work-point setting, vacuum, particle species, charge state, especially the electron cooler parameters such as electron density and radial distribution.

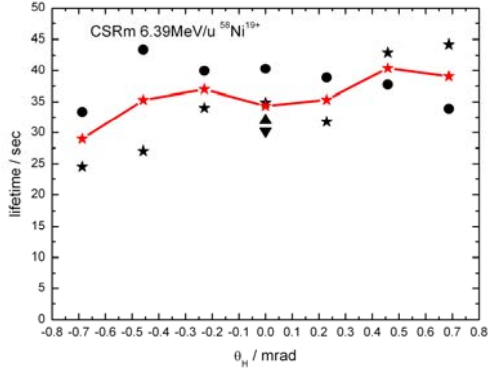

Figure 6a: The lifetime as a function of the related horizontal angle between ion and electron beams. The lifetime of ion beam in the CSRm was derived from the signal of DCCT.

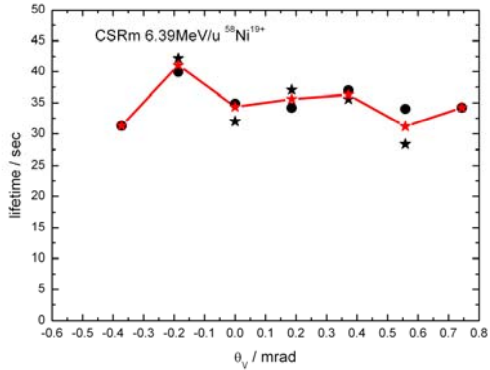

Figure 6b: The lifetime as a function of the related vertical angle between ion and electron beams.

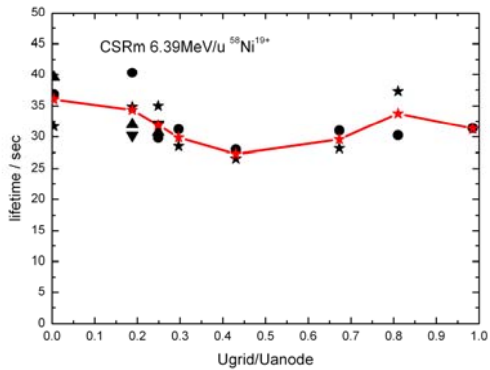

Figure 6c: The lifetime as a function of the profile of electron beam.

Capturing electron by the ion was the main beam loss way in the present of electron cooling and normal operation mode of storage ring. In this case, the temperature of electron was the important parameter. The possibility of capturing increased in the case of the lower related temperature between ions and electrons in the three degree of freedom, and the beam loss become serious, as a consequence the lifetime of ion beam became shorter. Fig 6a and 6b shows the lifetime of ion beam as a function of the related horizontal and vertical angle between ion and electron beams. From these results, the influence of angle on the ion beam lifetime was slight. When the electron moved in the cooler, the temperature was varied in the different situations. The maximal temperature increased when the angle existed, but the lowest temperature was not changed. The ions had their choices when the ion captured the free electron, as a result, no varying in the lifetime. The lifetime as a function of electron beam profile was shown in the Fig. 6c; the lifetime change is not obvious in the case of hollow electron beam as expected. The reason was the energy of electron beam was not perfectly matched in term of the varying of space charge in the different radial electron distribution.

# FREQUENCY SHIFT AND OSCILLATION

A related angle between ion and electron beams in the horizontal and vertical planes was intentionally created by the steering coils in the cooling section. In the vertical direction, the central frequency of ion beam shifted towards the lower side when the angle was bigger than a certain value. There was no obvious frequency shift in the case of smaller angles. In despite of the polarity of angle referenced to the zero, the tendency of central frequency shift was the same as demonstrated in The Fig. 8a. The frequency different is about $\Delta f/f=9.5\times 10^{-4}$. This can be interpreted as that an additional angle was created in the cooling section in the case of the fixed ion energy and dipole field of storage ring. This angle resulted in the increscent of ion beam path length in the vertical direction, in the other hands; the projection of the electron velocity in the ion beam orbit became smaller. The ion beam was drugged to the lower energy level, and the central frequency of ion beam became smaller compared with the situation of better alignment due to the both effects. In the horizontal plane, not only the central frequency of ion beam shifted towards the lower side, but also obvious oscillation appeared when the angle was bigger than a certain value. One conceivable explanation was that the high voltage of cooler was changed due to the electron beam hit in some place of cooler in the case of a bigger misalignment angle. The diameter of electron beam in the cooling section was about 59mm. The energy modulation of electron beam performed the similar behavior in the Schottky signal as exemplified in Fig. 9. The signal of modulation was uniform; the signal of oscillation was random. The different phenomena were observed in negative and positive angle in the horizontal plane. The central frequency shift and oscillation did not appear in the opposite angle, as shown in the Fig. 8b. The reasonable explanation was not found.

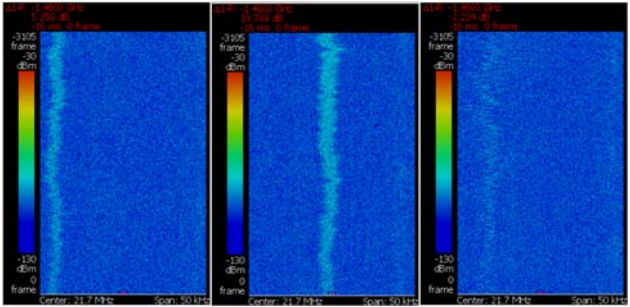

Figure 8a: The Schottky signals at different misalignment angle in the vertical plane (left: $\theta_V$=0.744mrad, centre: $\theta_V$=0mrad, right: $\theta_V$=-0.744mrad). The central frequency of ion beam moves to lower side in the case of a bigger misalignment in two directions. The frequency shift is not symmetrical.

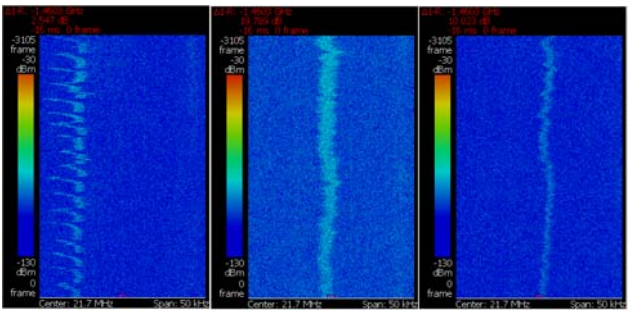

Figure 8b: The Schottky signals at different misalignment angle in the horizontal direction (left: $\theta_H$=0.687mrad, centre: $\theta_H$=0mrad, right: $\theta_H$=-0.687mrad). The central frequency of ion beam move to lower side, and the longitudinal oscillation was observed in the bigger misalignment in the horizontal plane.

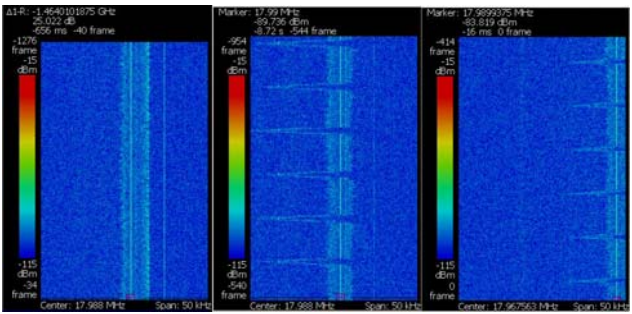

Figure 9: The Schottky signal during the modulation of electron beam energy. (left: no modulation, centre: 10V-100ms-1000ms, right: 20V-100ms-1000ms)

## PROGRESS, UPGRADE AND IMPROVEMENT IN CSR

- Add energy modulation system for CSRm cooler
- Improve the stability of power supply for CSRe dipoles
- Temperature control for 300kV cooler
- High voltage of CSRe cooler approach to 285kV
- 14 days continuous work at 285kV

## SUMMARY OF CSR STATUS

- The 35kV cooler can work at the lower energy (<1kV).
- The 300kV cooler can work at the higher energy (~285kV, 520MeV/u).
- The oscillation of ion beam was not caused by the instability of high voltage of cooler.
- Partial hollow electron beam is helpful to ion beam accumulation.
- A longitudinal oscillation signal was observed from Schottky probe during experiments.